%% file: main.tex
\documentclass[letterpaper, 10 pt, conference]{ieeeconf}
\IEEEoverridecommandlockouts
\let\labelindent\relax
\usepackage{enumitem}
\usepackage{balance}
\usepackage[font={small}]{caption}
\usepackage{subcaption}
\usepackage{array}
\usepackage{textcomp}
\usepackage{mathtools, nccmath}
\usepackage{graphicx}
\usepackage{amsfonts}
\usepackage{amsmath}
\usepackage{amssymb}
\usepackage{algorithm}
\usepackage{algorithmic}
\usepackage{hyperref}
\usepackage{tikz}
\usepackage{arydshln}
\usepackage{multirow}
\usepackage{bm}
\usepackage{epstopdf}
\usepackage{cite}
\usepackage{siunitx}

\DeclareMathOperator*{\argmin}{argmin} 

\newtheorem{remark}{Remark}

\begin{document}
\title{\LARGE \bf Safety-Critical Model Predictive Control with Discrete-Time \\ Control Barrier Function}
\author{Jun Zeng*, Bike Zhang* and Koushil Sreenath
    \thanks{* Authors have contributed equally and names are in alphabetical order.}
    \thanks{All authors are with the Department of Mechanical Engineering, University of California, Berkeley, CA, 94720, USA, \tt\small \{zengjunsjtu, bikezhang, koushils\}@berkeley.edu}
    \thanks{This work was partially supported through National Science Foundation Grant CMMI-1931853.}
    \thanks{Code is available at \url{https://github.com/HybridRobotics/MPC-CBF}}
}
\maketitle

\begin{abstract}
The optimal performance of robotic systems is usually achieved near the limit of state and input bounds. 
Model predictive control (MPC) is a prevalent strategy to handle these operational constraints, however, safety still remains an open challenge for MPC as it needs to guarantee that the system stays within an invariant set.
In order to obtain safe optimal performance in the context of set invariance, we present a safety-critical model predictive control strategy utilizing discrete-time control barrier functions (CBFs), which guarantees system safety and accomplishes optimal performance via model predictive control.
We analyze the feasibility and the stability properties of our control design.
We verify the properties of our method on a 2D double integrator model for obstacle avoidance.
We also validate the algorithm numerically using a competitive car racing example, where the ego car is able to overtake other racing cars.
\end{abstract}

\IEEEpeerreviewmaketitle
\input{introduction.tex}
\input{background.tex}
\input{formulation.tex}

\input{example.tex}
\input{conclusion.tex}
\vspace{-1mm}
\bibliographystyle{IEEEtran}
\bibliography{references}{}
\balance
\input{appendix.tex}
\end{document}

%% file: introduction.tex
\section{Introduction}
\label{sec:introduction}
\subsection{Motivation}
Safety-critical optimal control and planning is one of the fundamental problems in robotic applications.
In order to ensure the safety of robotic systems while achieving optimal performance, the tight coupling between potentially conflicting control objectives and safety criteria is considered in an optimization problem.
Some recent work formulates this problem using control barrier functions, but only using current state information without prediction, see \cite{ames2014control, agrawal2017discrete}, which yields a greedy control policy.
Model predictive control can give a less greedy policy, as it takes future state information into account. However, the safety criteria in a predictive control framework is usually enforced as distance constraints defined under Euclidean norms, such as the distance between the robot and obstacles being larger than a safety margin.
This distance constraint will not confine the optimization until the reachable set along the horizon intersects with the obstacles.
In other words, the robot will not take actions to avoid the obstacles until it is close to them. One way to solve this problem is to use a larger horizon, but that will increase the computational complexity in the optimization.   

\begin{figure}
    \centering
    \begin{subfigure}[t]{0.99\linewidth}
        \centering
        \includegraphics[width=1\linewidth]{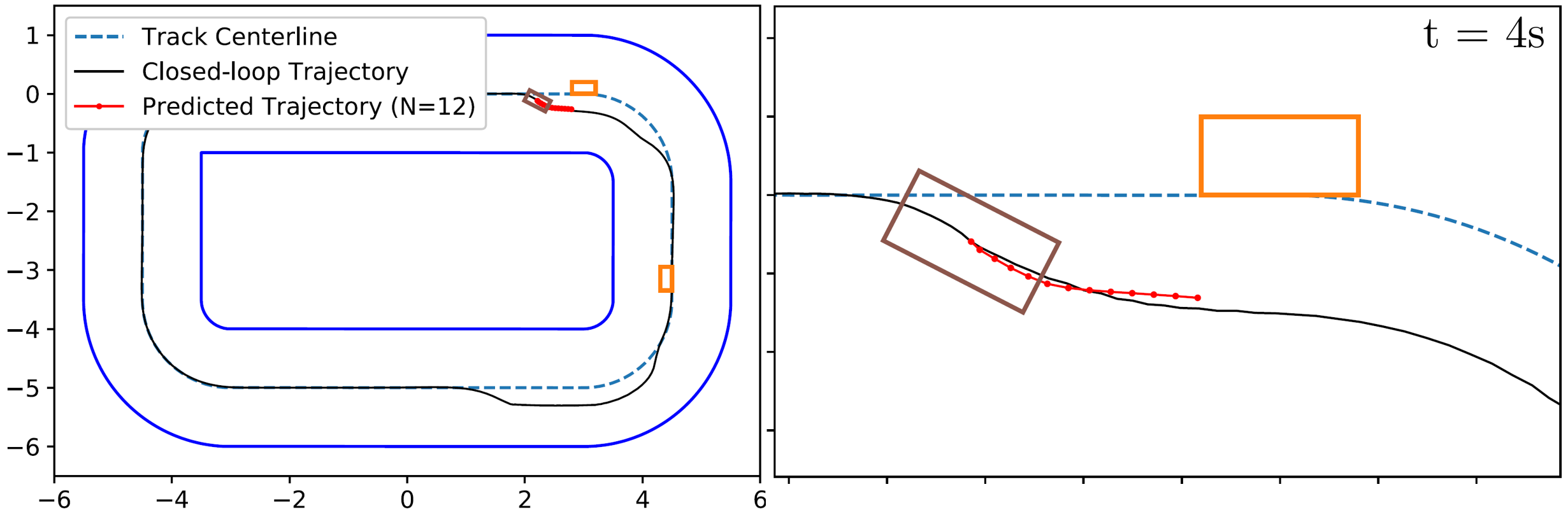}
        \caption{Snapshot of first overtake [whole track (left) vs. zoom-in (right)]}
        \label{fig:cover_first}
    \end{subfigure}
    \begin{subfigure}[t]{0.99\linewidth}
        \centering
        \includegraphics[width=1\linewidth]{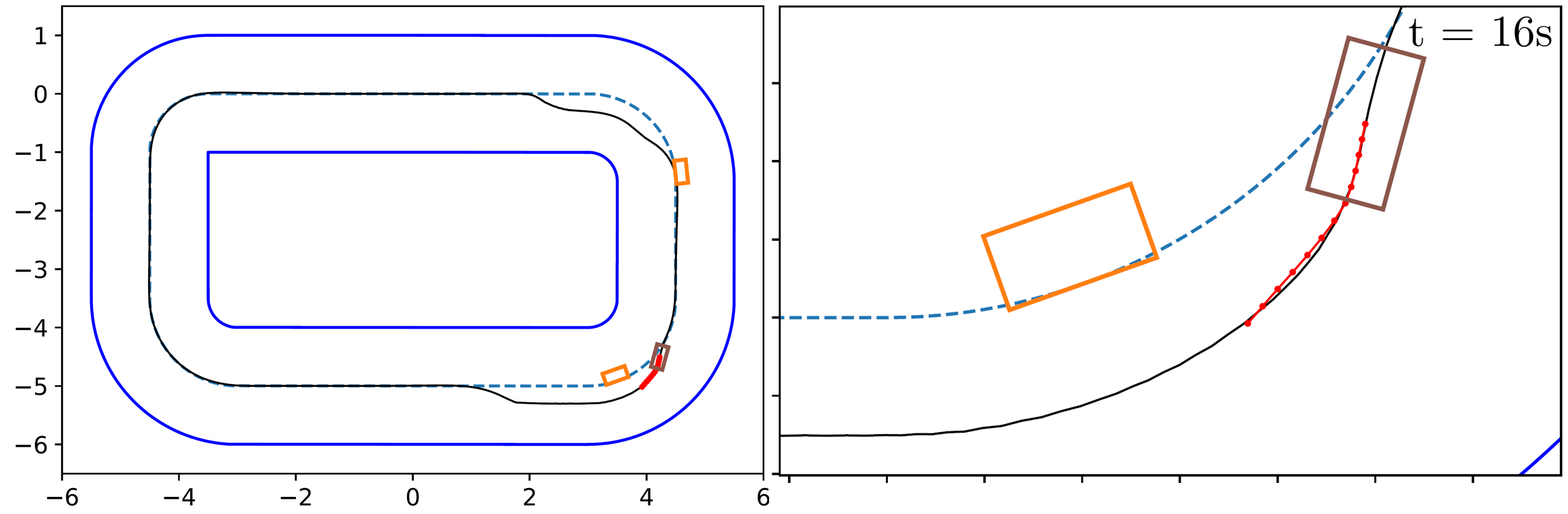}
        \caption{Snapshot of second overtake [whole track (left) vs. zoom-in (right)]}
        \label{fig:cover_second}
    \end{subfigure}
    \caption{The proposed safety-critical model predictive control is applied to a competitive car racing example. Snapshots of the ego car (brown) are shown overtaking other racing cars (orange) from both right and left sides while maintaining a specified target speed. The closed-loop trajectory and predicted open-loop trajectory are colored in black and red respectively, and the blue solid lines depict the boundary of racing track.}
    \label{fig:cover}
\end{figure}

We address this challenge above by directly unifying model predictive control with discrete-time control barrier functions together into one optimization problem.
This results in a safety-critical model predictive control formulation, called MPC-CBF in this paper.
In this formulation, the CBF constraints could enforce the system to avoid obstacles even when the reachable set along the horizon is far away from the obstacles. 
We validate this control design using a 2D double integrator for obstacle avoidance, and also demonstrate that this method enables a racing car to safely compete with other cars in a racing competition, shown in Fig. \ref{fig:cover}.

\subsection{Related Work}

\subsubsection{Model Predictive Control}
MPC is widely used for robotic systems, such as robotic manipulation and locomotion \cite{hogan2020reactive, scianca2020mpc}, to achieve optimal performance while satisfying different constraints.
One of the most important criteria to deploy robots for real-world tasks is safety. 
There is some existing work about model predictive control considering system safety \cite{son2019safety, rosolia2020multi, rosolia2020unified}.
The safety criteria in the context of MPC is usually formulated as constraints in an optimization problem \cite{yoon2009model, frasch2013auto}, such as obstacle constraints and actuation limits. 
One concrete scenario regarding safety criteria for robots is obstacle avoidance.
The majority of literature focuses on collision avoidance using simplified models, and considers distance constraints with various Euclidean norms \cite{turri2013linear,rosolia2016autonomous,zhang2020optimization}, which we call MPC-DC in this paper.

However, these obstacle avoidance constraints under Euclidean norms will not confine the robot's movement unless the robot is relatively close to the obstacles.
To make the robot take actions to avoid obstacles even far away from it, we usually need a larger horizon which increases the computational time in the optimization.
This encourages us to formulate a new type of model predictive control, which can guarantee safety in the context of set invariance with CBF constraints confining the robot's movement during the optimization. Recently in \cite{grandia2020nonlinear}, model predictive control is introduced with control Lyapunov functions to ensure stability.

\subsubsection{Control Barrier Functions}
CBFs have recently been introduced as a promising way to ensure set invariance by considering the system dynamics and implying forward invariance of the safe set.
Furthermore, a safety-critical control design for continuous-time systems was proposed by unifying a control Lyapunov function (CLF) and a control barrier function through a quadratic program (CLF-CBF-QP) \cite{ames2019control}.
This method could be deployed as a real-time optimization-based controller with safety-critical constraints, shown in \cite{wu2015safety, ames2014control, nguyen20163d}.
Besides the continuous-time domain, the formulation of CBFs was generalized into discrete-time systems (DCLF-DCBF) in \cite{agrawal2017discrete}.

\subsubsection{Model Predictive Control with Control Barrier Functions}
There is some existing work that tries to combine the advantages of MPC and CBFs. Barrier functions have been used in MPC in \cite{wills2004barrier}, which converts constraints to cost but not related to safety-critical control. In \cite{son2019safety}, they use continuous-time CBFs as constraints inside a discrete-time MPC.
MPC and CBFs are organized as a high-level planner and a low-level tracker in \cite{rosolia2020multi}. This method treats MPC and CBFs separately at different levels.

Inspired by the previous work of model predictive control and control barrier functions, DCLF-DCBF can be improved by taking future state prediction into account, yielding a better control policy. This motivates us to investigate the control design of predictive control under the constraints imposed by CBFs.
In this paper, we focus on the discrete-time formulation of control barrier functions applied to model predictive control, which encodes the safety obtained from discrete-time CBFs in MPC.

\subsection{Contribution}
The contributions of this paper are as follows.
\begin{itemize}
    \item We present a MPC-CBF control design for safety-critical tasks, where the safety-critical constraints are enforced by discrete-time control barrier functions.
    \item We analyze the stability of our control design, and qualitatively discuss the feasibility in terms of set intersections between reachable sets of MPC and safe sets enforced by CBF constraints along the horizon.
    \item Our proposed method is shown to outperform both MPC-DC and DCLF-DCBF. It enables prediction capability to DCLF-DCBF for performance improvement, and it also guarantees safety via discrete-time CBF constraints in the context of set invariance. 
    \item We verify the properties of our control design using a 2D double integrator for obstacle avoidance. Our algorithm is generally applicable and also validated in a more complex scenario, where MPC-CBF enables a car racing on a track while safely overtaking other cars.
\end{itemize}

\subsection{Paper Structure}
This paper is organized as follows:
in Sec. \ref{sec:background}, we present the background of model predictive control and control barrier functions.
In Sec. \ref{sec:control-design}, we introduce the safety-critical model predictive control design using discrete-time control barrier functions (MPC-CBF).
The analysis of stability and feasibility properties is presented and the relations with DCLF-DCBF and MPC-DC are also discussed.
To validate the control design and verify the properties of our formulation, a 2D double integrator for obstacle avoidance and a car racing competition example are demonstrated in Sec. \ref{sec:example}.
Sec. \ref{sec:conclusion} provides concluding remarks.

%% file: background.tex
\section{Background}
\label{sec:background}

Our proposed safety-critical model predictive control design builds on model predictive control and control barrier functions. We now present necessary preliminaries. 

\subsection{Model Predictive Control}
Consider the problem of regulating to the origin of a discrete-time control system described by,
\begin{equation}
    \mathbf{x}_{t+1} = f(\mathbf{x}_t, \mathbf{u}_t), \label{eq:discrete-time-dynamics}
\end{equation}
where $\mathbf{x}_t \in \mathcal{X} \subset \mathbb{R}^n$ represents the state of the system at time step $t \in \mathbb{Z}^{+}$, $\mathbf{u}_t \in \mathcal{U} \subset \mathbb{R}^m$ is the control input, and $f$ is locally Lipschitz. 

Assume that a full measurement or estimate of the state $\mathbf{x}_t$ is available at the current time step $t$.
Then a finite-time optimal control problem is solved at time step $t$.
When there are safety criteria, such as obstacle avoidance, the obstacles are usually formulated using distance constraints.
The finite-time optimal control formulation is shown in \eqref{eq:mpc-ftoc}.
\noindent\rule{\columnwidth}{0.4pt}
\textbf{MPC-DC:}
\begin{subequations}
\label{eq:mpc-ftoc}
\begin{align}
    J_{t}^{*}(\mathbf{x}_t) =  \min_{\mathbf{u}_{t:t{+}N{-}1|t}} p(\mathbf{x}_{t+N|t}){+}\sum_{k=0}^{N-1} & q(\mathbf{x}_{t+k|t},\mathbf{u}_{t+k|t}) \label{eq:mpc-cost}\\
    \text{s.t.} \quad 
    \mathbf{x}_{t+k+1|t} = f(\mathbf{x}_{t+k|t}, \mathbf{u}_{t+k|t}), & \ k = 0,...,N{-}1  \label{eq:mpc-system-dynamics}\\
    \mathbf{x}_{t+k|t} \in \mathcal{X}, \mathbf{u}_{t+k|t} \in \mathcal{U}, & \ k = 0,...,N{-}1 \label{eq:mpc-state-input-constraint}\\
    \mathbf{x}_{t|t} = \mathbf{x}_t, \label{eq:mpc-current-state} \\
    \mathbf{x}_{t+N|t} \in \mathcal{X}_f, & \label{eq:mpc-terminal-set}\\
    g(\mathbf{x}_{t+k|t}) \geq 0, &\ k = 0,...,N{-}1. \label{eq:mpc-distance-constraint}
\end{align}
\end{subequations}
\noindent\rule{\columnwidth}{0.4pt}
Here $\mathbf{x}_{t+k|t}$ denotes the state vector at time step $t$~+~$k$ predicted at time step $t$ obtained by starting from the current state $\mathbf{x}_t$ \eqref{eq:mpc-current-state}, and applying the input sequence $\mathbf{u}_{t:t+N-1|t}$ to the system dynamics \eqref{eq:mpc-system-dynamics}. 
In \eqref{eq:mpc-cost}, the terms $q(\mathbf{x}_{t+k|t},\mathbf{u}_{t+k|t})$ and $p(\mathbf{x}_{t+N|t})$ are referred to as stage cost and terminal cost respectively, and N is the time horizon. 
The state and input constraints are given by \eqref{eq:mpc-state-input-constraint}, and distance constraints for safety criteria are represented by function $g$, in \eqref{eq:mpc-distance-constraint}, which could be defined under various Euclidean norms.
The terminal constraint is enforced in \eqref{eq:mpc-terminal-set}.

Let $\mathbf{u}_{t:t{+}N{-}1|t}^{*} = \{\mathbf{u}_{t|t}^{*},...,\mathbf{u}_{t{+}N{-}1|t}^{*}\}$ be the optimal solution of \eqref{eq:mpc-ftoc} at time step $t$. The resulting optimized trajectory using $\mathbf{u}_{t:t{+}N{-}1|t}^{*}$ is referred as an open-loop trajectory.
Then, the first element of $\mathbf{u}_{t:t{+}N{-}1|t}^{*}$ is applied to system \eqref{eq:discrete-time-dynamics}. This feedback control law is given below,
\begin{equation}
    \mathbf{u}(t) = \mathbf{u}_{t|t}^{*}(\mathbf{x}_t). \label{eq:mpc-law}
\end{equation}
The finite-time optimal control problem \eqref{eq:mpc-ftoc} is repeated at next time step $t+1$, based on the new estimated state $\mathbf{x}_{t+1|t+1} = \mathbf{x}_{t+1}$. It yields the model predictive control strategy.
The resulting trajectory using \eqref{eq:mpc-law} is referred as a closed-loop trajectory.
More details can be referred to in \cite{borrelli2017predictive}.

\subsection{Control Barrier Functions}
We now present discrete-time control barrier functions that will be used together with model predictive control for our control design, which will be introduced in Sec. \ref{sec:control-design}.

For safety-critical control, we consider a set $\mathcal{C}$ defined as the superlevel set of a continuously differentiable function $h: \mathcal{X} \subset \mathbb{R}^n \rightarrow \mathbb{R}$:
\begin{equation}
    \mathcal{C} = \{ \mathbf{x} \in \mathcal{X} \subset \mathbb{R}^n: h(\mathbf{x}) \geq 0\}. \label{eq:cbf-safeset}
\end{equation}
Throughout this paper, we refer to $\mathcal{C}$ as a safe set. The function $h$ is a control barrier function (CBF)~\cite{ames2014control} if {\small $\dfrac{\partial h}{\partial \mathbf{x}} \neq 0$} for all $\mathbf{x} \in \partial \mathcal{C}$  and there exists an extended class $\mathcal{K}_{\infty}$ function $\gamma$ such that for the control system \eqref{eq:discrete-time-dynamics}, $h$ satisfies
\begin{equation}
    \exists ~\mathbf{u} ~\text{s.t.} ~\dot{h}(\mathbf{x}, \mathbf{u}) \geq -\gamma(h(\mathbf{x})), ~\gamma \in \mathcal{K}_{\infty}. \label{eq:cbf-original-definition}
\end{equation}
This condition can be extended to the discrete-time domain which is shown as follows
\begin{equation}
    \Delta h(\mathbf{x}_k, \mathbf{u}_k) \geq -\gamma h(\mathbf{x}_k), ~ 0 < \gamma \leq 1, \label{eq:cbf-definition}
\end{equation}
where $\Delta h(\mathbf{x}_k, \mathbf{u}_k) := h(\mathbf{x}_{k+1})-h(\mathbf{x}_k)$.
Satisfying constraint \eqref{eq:cbf-definition}, we have $h(\mathbf{x}_{k+1}) \geq (1-\gamma) h(\mathbf{x}_k)$, i.e., the lower bound of control barrier function $h(x)$ decreases exponentially with the rate $1-\gamma$.

\begin{remark}
Note that in \eqref{eq:cbf-definition}, we defined $\gamma$ as a scalar instead of a $\mathcal{K}_{\infty}$ function as in \eqref{eq:cbf-original-definition}.
Generally, for the discrete-time domain, $\gamma$ could also be considered as a class $\mathcal{K}$ function that also satisfies $0 < \gamma(h(\mathbf{x})) \leq h(\mathbf{x})$ for any $h(\mathbf{x})$. 
However, we will continue to use the scalar form $\gamma$ in this paper to simplify the notations for further discussions.
\end{remark}

Besides the system safety, we are also interested in stabilizing the system with a feedback control law $\mathbf{u}$ under a control Lyapunov function $V$, i.e.,
\begin{equation}
    \exists ~\mathbf{u} ~\text{s.t.} ~\dot{V}(\mathbf{x}, \mathbf{u}) \leq -\alpha(V(\mathbf{x})), ~\alpha \in \mathcal{K}. \label{eq:clf-original-definition}
\end{equation}
We can also generalize it to the discrete-time domain,
\begin{equation}
    \Delta V(\mathbf{x}_k, \mathbf{u}_k) \leq -\alpha V(\mathbf{x}_k), ~0 < \alpha \leq 1,
\end{equation}
where $\Delta V(\mathbf{x}_k, \mathbf{u}_k) := V(\mathbf{x}_{k+1}) - V(\mathbf{x}_k)$. Similarly as above, the upper bound of control Lyapunov function decreases exponentially with the rate $1-\alpha$.

The discrete-time control Lyapunov function and control barrier function can be unified into one optimization program (DCLF-DCBF), which achieves the control objective and guarantees system safety. This formulation was first introduced in \cite{agrawal2017discrete} and is presented as follows
\noindent\rule{\columnwidth}{0.4pt}
\textbf{DCLF-DCBF:}
\begin{subequations}
\label{eq:CLF-CBF-discrete}
\begin{align}
    \mathbf{u}_k^* = \argmin_{(\mathbf{u}_k, \delta) \in \mathbb{R}^{m+1}} \quad & \mathbf{u}_k^T H(\mathbf{x}) \mathbf{u}_k + l \cdot \delta^2 \label{eq:CLF-CBF-discrete-cost}\\
    \quad & \Delta V(\mathbf{x}_k, \mathbf{u}_k) + \alpha V(\mathbf{x}_k) \leq \delta, \label{eq:CLF-CBF-discrete-CLF}\\
    \quad & \Delta h(\mathbf{x}_k, \mathbf{u}_k) + \gamma h(\mathbf{x}_k) \geq 0, \label{eq:CLF-CBF-discrete-CBF}\\
    \quad & \mathbf{u}_k \in \mathcal{U},
\end{align}
\end{subequations}
\noindent\rule{\columnwidth}{0.4pt}
where $H(\mathbf{x})$ is a positive definite matrix, that is pointwise differentiable in $\mathbf{x}$. $l$ is positive, and $\delta \geq 0$ is a slack variable that allows the control Lyapunov function to grow when the CLF and CBF constraints are conflicting. The safe set $\mathcal{C}$ in \eqref{eq:cbf-safeset} is invariant along the trajectories of the discrete-time system with controller \eqref{eq:CLF-CBF-discrete} if $h(\mathbf{x}_0)$ $\geq$ 0 and 0 $<$ $\gamma$ $\leq$ 1.

%% file: formulation.tex
\section{Control Design}
\label{sec:control-design}
After presenting a background of model predictive control and control barrier functions, we formulate the safety-critical model predictive control logic in this section.
\subsection{Formulation}
\label{subsec:formulation}
Consider the problem of regulating to a target state for the discrete-time system \eqref{eq:discrete-time-dynamics} while ensuring safety in the context of set invariance.
The proposed control logic MPC-CBF solves the following constrained finite-time optimal control problem with horizon $N$ at each time step $t$
\noindent\rule{\columnwidth}{0.4pt}
\textbf{MPC-CBF:}
\begin{subequations}
\label{eq:mpc-cbf}
\begin{align}
    J_{t}^{*}(\mathbf{x}_t){=}\min_{\mathbf{u}_{t:t{+}N{-}1|t}} p(\mathbf{x}_{t+N|t}){+}\sum_{k=0}^{N-1}&q(\mathbf{x}_{t+k|t}, \mathbf{u}_{t+k|t}) \label{eq:mpc-cbf-cost}\\
    \text{s.t.} \quad 
    \mathbf{x}_{t+k+1|t} = f(\mathbf{x}_{t+k|t}, \mathbf{u}_{t+k|t}), &\ k = 0,...,N{-}1\label{eq:mpc-cbf-dynamics} \\
    \mathbf{x}_{t+k|t} \in \mathcal{X}, \mathbf{u}_{t+k|t} \in \mathcal{U}, &\ k = 0,...,N{-}1 \label{eq:mpc-cbf-constraint}\\
    \mathbf{x}_{t|t} = \mathbf{x}_t, & \label{eq:mpc-cbf-initial-condition}\\
    \mathbf{x}_{t+N|t} \in \mathcal{X}_f, & \label{eq:mpc-cbf-terminal-set}\\
    \Delta h (\mathbf{x}_{t+k|t}, \mathbf{u}_{t+k|t}) \geq -\gamma h(\mathbf{x}_{t+k|t}), &\ k = 0,...,N{-}1 \label{eq:mpc-cbf-cbf}
\end{align}
\end{subequations}
\noindent\rule{\columnwidth}{0.4pt}
\noindent
where \eqref{eq:mpc-cbf-initial-condition} represents the initial condition constraint, \eqref{eq:mpc-cbf-dynamics} describes the system dynamics, and \eqref{eq:mpc-cbf-constraint} shows the state/input constraints along the horizon.
The terminal set constraint is imposed in \eqref{eq:mpc-terminal-set}.
The CBF constraints in \eqref{eq:mpc-cbf-cbf} are designed to guarantee the forward invariance of the safe set $\mathcal{C}$ associated with the discrete-time control barrier function, where $\Delta h$ is as introduced in \eqref{eq:cbf-definition}. Here we have
\begin{equation*}
    \Delta h (\mathbf{x}_{t+k|t}, \mathbf{u}_{t+k|t}) = h(\mathbf{x}_{t+k+1|t}) - h(\mathbf{x}_{t+k|t}).
\end{equation*}

The optimal solution to \eqref{eq:mpc-cbf} at time $t$ is a sequence of inputs as $\mathbf{u}_{t:t+N-1|t}^{*} = [\mathbf{u}_{t|t}^*,...,\mathbf{u}_{t+N-1|t}^*]$. Then, the first element of the optimizer vector is applied
\begin{equation}
\mathbf{u}(t) = \mathbf{u}_{t|t}^{*}(\mathbf{x}_t). \label{eq:mpc-cbf-law}
\end{equation}
This constrained finite-time optimal control problem \eqref{eq:mpc-cbf} is repeated at time step $t+1$, based on the new state $\mathbf{x}_{t+1|t+1}$, yielding a receding horizon control strategy: safety-critical model predictive control (MPC-CBF).

The dynamics constraint in \eqref{eq:mpc-cbf-dynamics} could be linear if we have a linear system, and \eqref{eq:mpc-cbf-constraint} could also be linear if $\mathcal{X}$ and $\mathcal{U}$ are defined as polytopes in the state and input space, respectively.
The discrete-time control barrier functions constraints in \eqref{eq:mpc-cbf-cbf} are generally non-convex unless the CBFs are linear.
This makes the whole optimization in \eqref{eq:mpc-cbf} generally become a nonlinear programming problem (NLP).

\subsection{Stability}
In DCLF-DCBF, control Lyapunov functions are introduced as optimization constraints \eqref{eq:CLF-CBF-discrete-CLF} with corresponding slack variable as additional term in the cost function \eqref{eq:CLF-CBF-discrete-cost}.
This allows for the achievement of control objectives represented by CLFs and unifies the formulation under one optimization with CBFs.
In our MPC-CBF control design, we have the terminal cost $p(\mathbf{x}_{t+N|t})$ in \eqref{eq:mpc-cbf-cost} as a control Lyapunov function, which can be used to guarantee the stability of the closed-loop system by satisfying mild assumptions in the context of linear systems \cite[Thm. 12.2]{borrelli2017predictive}.
In general, a rigorous proof of nonlinear MPC stability as well as MPC-CBF still remains an open challenge. More detailed analysis could be found in \cite{allgower2012nonlinear}.

\subsection{Feasibility}
\label{subsec:feasibility}
Recursive feasibility is generally not guaranteed for MPC-DC defined in \eqref{eq:mpc-ftoc} \cite{borrelli2017predictive, lars2011nonlinear} and other general NLP \cite[Chap. 11]{boyd2004convex}. In this paper, we qualitatively analyze the feasibility problem of MPC-CBF based on set analysis.
Since the optimization \eqref{eq:mpc-cbf} could be a NLP, we are interested in finding under which circumstances this optimization becomes feasible, i.e., the feasible set under the constraints \eqref{eq:mpc-cbf-dynamics}-\eqref{eq:mpc-cbf-cbf} is not empty.
\begin{figure}
    \centering
    \includegraphics[width=\linewidth]{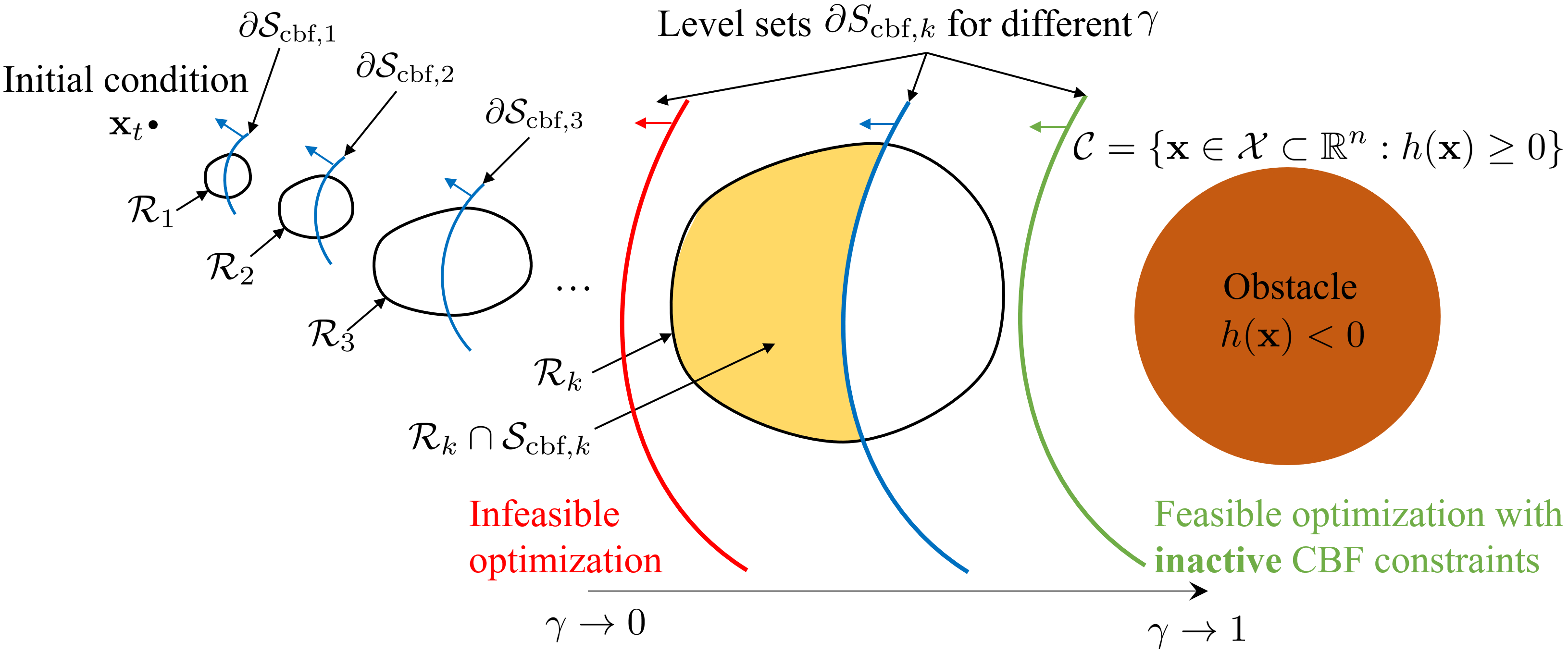}
    \caption{Feasibility of MPC-CBF. The reachable set $\mathcal{R}_k$ propagates along the horizon from the initial condition $x_t$. For horizon step $k$ in the open-loop, the level sets $\partial \mathcal{S}_{\text{cbf},k}$ are shown in different colors with three choices of $\gamma$ and each corresponding $\mathcal{S}_{\text{cbf},k}$ lies on the left hand side of level sets, indicated by the arrows in different colors.}
    \label{fig:mpc-cbf-feasibility}
\end{figure}

Given the current state $\mathbf{x}_t$ in \eqref{eq:mpc-cbf-initial-condition}, the reachable set at horizon step $k$ is defined as a reachable region in the state space, satisfying system dynamics in \eqref{eq:mpc-cbf-dynamics}, input/state constraints in \eqref{eq:mpc-cbf-constraint} and initial condition in \eqref{eq:mpc-cbf-initial-condition}. This reachable set $\mathcal{R}_k$ is defined as follows for the horizon step $k$:
\begin{equation}
\begin{split}
    \mathcal{R}_k = \{\mathbf{x}_{t+k|t} \in \mathcal{X}: \forall i = 0,...,k-1, \\
    \mathbf{x}_{t+i+1|t} = f(\mathbf{x}_{t+i|t}, \mathbf{u}_{t+i|t}), \\ 
    \mathbf{x}_{t+i|t} \in \mathcal{X}, \mathbf{u}_{t+i|t} \in \mathcal{U}, \mathbf{x}_{t|t} = \mathbf{x}_t\}.
\end{split}
\end{equation}

We also define the set of state space satisfying the CBF constraints in \eqref{eq:mpc-cbf-cbf} and initial condition in \eqref{eq:mpc-cbf-initial-condition} as 
\begin{equation}
\begin{split}
    \mathcal{S}_{\text{cbf},k} = \left\{ \right. & \mathbf{x} \in \mathcal{X} : \\
    & \left. h(\mathbf{x}){-}h(\mathbf{x}_{t+k-1|t}) {\geq} {-}\gamma h (\mathbf{x}_{t+k-1|t}) \right\},
\end{split}
\end{equation}
where $\mathcal{S}_{\text{cbf},k}$ describes superlevel sets of $h$ satisfying the control barrier function constraints \eqref{eq:mpc-cbf-cbf} at each time step along the open-loop trajectory.
$\mathcal{S}_{\text{cbf},k}$ also depends on the value of optimal value $\mathbf{x}_{t+k-1|t}$, which depends on the states and the inputs of previous nodes before the index $k-1$.

We illustrate $\mathcal{R}_k$ and $\mathcal{S}_{\text{cbf},k}$ in the state space shown in Fig. \ref{fig:mpc-cbf-feasibility}, given the initial condition $\mathbf{x}_t$.
Then, the feasibility of the optimization in \eqref{eq:mpc-cbf} turns out to be whether the intersection between the feasible set at each horizon step, $\mathcal{R}_k$, and the superlevel set of $h(\mathbf{x})$ satisfying the CBF constraints, $\mathcal{S}_{\text{cbf},k}$, is nonempty for all $k = 1,...,N$.
\begin{remark}
Note that $\mathcal{S}_{\text{cbf},k}$ is not empty when $h$ is a valid control barrier function.  Furthermore, 
$\mathcal{R}_k$ is also guaranteed to be nonempty, if we choose $\mathcal{X}$ as a control invariant set as discussed in \cite[Thm. 11.2]{borrelli2017predictive}.
\end{remark}

In order to better understand this problem, we illustrate the level sets of control barrier function constraints as
\begin{equation*}
    \partial \mathcal{S}_{\text{cbf},k} = \{x \in \mathcal{X}: h(\mathbf{x}) = (1-\gamma) h(\mathbf{x}_{t+k-1|t})\},
\end{equation*}
with several choices of $\gamma$.
The superlevel set $\mathcal{S}_{\text{cbf},k}$ are the regions illustrated on the left hand side of these level sets, with an example where the robot approaches the obstacle from the left to the right, shown in Fig. \ref{fig:mpc-cbf-feasibility}.
We can see that, if $\gamma$ becomes relatively small, $\mathcal{S}_{\text{cbf},k}$ will be smaller.
In this case, the system tends to be safer as the decar-rate of CBF becomes smaller, but the intersection between $\mathcal{R}_k$ and $\mathcal{S}_{\text{cbf},k}$ might be infeasible if $\gamma$ becomes too small.
When the $\gamma$ becomes larger, the region of $\mathcal{S}_{\text{cbf},k}$ in the state space will be increased.
This will make the optimization more likely to be feasible, however, the CBF constraints might not be active during the optimization, if $\gamma$ is too large.
In this case, $\mathcal{R}_k$ will become a proper subset of $\mathcal{S}_{\text{cbf},k}$.

\begin{figure*}
    \centering
    \includegraphics[width = 0.95\linewidth]{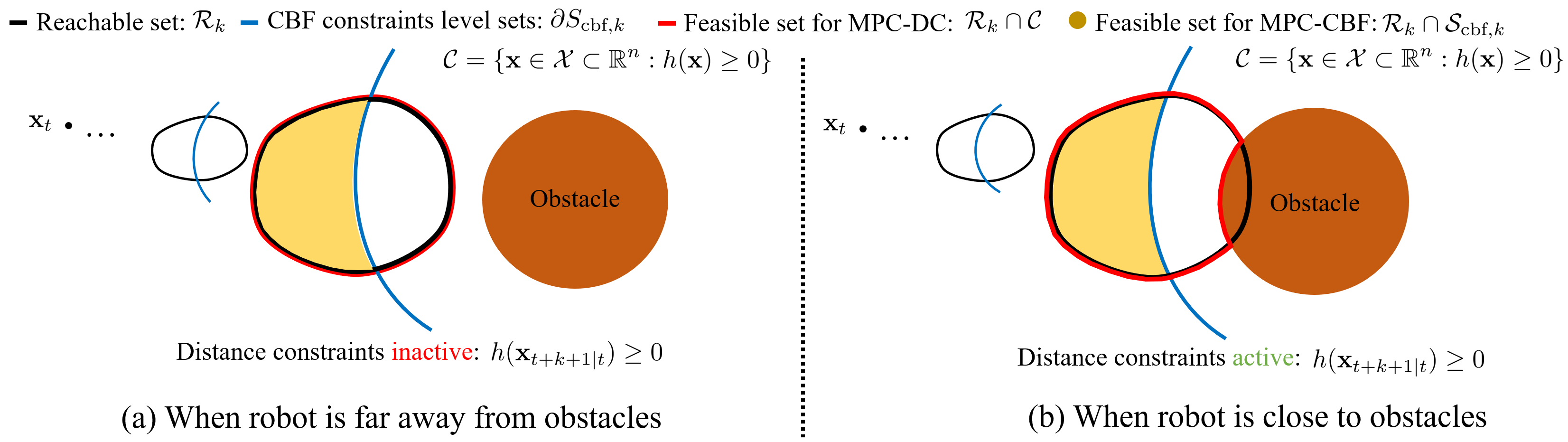}
    \caption{
    For MPC-CBF, the feasible set at each horizon step $k$ is the intersection between $\mathcal{R}_k$ and $\mathcal{S}_{\text{cbf},k}$, colored in yellow.
    For MPC-DC, the feasible set at each horizon step is the intersection between $\mathcal{R}_k$ and $\mathcal{C}$, colored with borders in red.
    We clearly see that MPC-CBF is safer than MPC-DC with smaller set invariance.
    Moreover, distance constraints might be inactive when the robot is still far away from obstacles, illustrated in (a), where $\mathcal{R}_k$ is a subset of $\mathcal{C}$.
    CBF constraints could still confine the robot's reachable set $\mathcal{R}_k$ with appropriate choices of $\gamma$ even when the robot is far away from obstacles, shown in (a) and (b).
    }
    \label{fig:mpc-cbf-vs-mpc-dc-constraint-activation}
\end{figure*}

\begin{remark}
When $\gamma$ becomes relatively small, the MPC-CBF controller makes a smaller subset of the safe set $\mathcal{C}$ in \eqref{eq:cbf-safeset} invariant, and thus is more safer, but this might also make the optimization infeasible.
A larger $\gamma$ will make the optimization more likely to be feasible, but the CBF constraints might not be active during the optimization.
We expect that $\gamma$ is chosen appropriately, such that the intersection between these two sets will not be empty and becomes a proper subset of $\mathcal{R}_k$.
This leads to a tradeoff between safety and feasibility in terms of the choice of $\gamma$.
However, it still remains an open challenge for how to automatically choose the $\gamma$.
\end{remark}

\begin{remark}
Given $\mathbf{x}_t$, $\mathcal{X}$, $\mathcal{U}$, we could pick a value of $\gamma$ to find a tradeoff between safety and feasibility.
However, when the system evolves, this $\gamma$ might no longer satisfy our safety demand or guarantee the optimization feasibility.
Therefore, for a given fixed $\gamma$, we generally only have pointwise feasibility and a persistently feasible formulation is still an open problem and is part of future work. 
\end{remark}

\subsection{Relation with DCLF-DCBF}
\label{subsec:relation-dclf-dcbf}
When $N=1$, the formulation in \eqref{eq:mpc-cbf} could be simplified as an optimization over one step system input $\mathbf{u}_k^*$
\begin{subequations}
\label{eq:mpc-cbf-one-horizon}
\begin{align}
    \mathbf{u}_k^* = \argmin_{\mathbf{u}_k \in \mathbb{R}^m} \quad & p(f(\mathbf{x}_k, \mathbf{u}_k)) + q(\mathbf{x}_k, \mathbf{u}_k) \label{eq:mpc-cbf-one-horizon-cost} \\
    \quad & \Delta h(\mathbf{x}_k, \mathbf{u}_k) + \gamma h(\mathbf{x}_k) \geq 0, \label{eq:mpc-cbf-one-horizon-CBF} \\
    \quad & \mathbf{u}_k \in \mathcal{U},
\end{align}
\end{subequations}
where $p(f(\mathbf{x}_k, \mathbf{u}_k))$ and $q(\mathbf{x}_k, \mathbf{u}_k)$ are the terminal cost and stage cost which we have seen previously in \eqref{eq:mpc-cbf-cost}.
The optimization \eqref{eq:mpc-cbf-one-horizon} is similar to the DCLF-DCBF formulation \eqref{eq:CLF-CBF-discrete}. 
The stage cost $q(\mathbf{x}_k, \mathbf{u}_k)$ minimizes the system input, similar as $\mathbf{u}_k^T H(\mathbf{x}) \mathbf{u}_k$ in \eqref{eq:CLF-CBF-discrete-cost}.
The terminal cost minimizes the control Lyapunov function $p(f(\mathbf{x}_k, \mathbf{u}_k))$ in \eqref{eq:mpc-cbf-one-horizon-cost}, instead of using CLF constraints as \eqref{eq:CLF-CBF-discrete-CLF}.
As we no longer use the CLF constraint, we do not need a slack variable, such as $\delta$ in \eqref{eq:CLF-CBF-discrete-cost}, to guarantee the feasibility. 
\begin{remark}
As the CLF constraints in \eqref{eq:CLF-CBF-discrete-CLF} are transferred into the cost function in \eqref{eq:mpc-cbf-one-horizon} with $N=1$, the formulation in \eqref{eq:mpc-cbf-one-horizon} becomes similar to DCLF-DCBF.
To sum up, our MPC-CBF formulation operates in a similar manner of DCLF-DCBF with prediction $N = 1$.
\end{remark}

\subsection{Relation with MPC-DC}
\label{subsec:relation-mpc-dc}
When $\gamma$ approaches its upper bound of 1, the CBF constraints in \eqref{eq:mpc-cbf-cbf} becomes,
\begin{equation*}
    h (\mathbf{x}_{t+k+1|t}) \geq 0.
\end{equation*}
If $g(\mathbf{x})$ in \eqref{eq:mpc-distance-constraint} and $h(\mathbf{x})$ are the same, these CBF constraints are almost the same as distance constraints defined in \eqref{eq:mpc-distance-constraint} except for one horizon step difference. In other words, the CBF constraints are at the next predicted horizon step instead of the current horizon step.
Moreover, the feasible set at each horizon step $k$ in MPC-DC formulation becomes $\mathcal{R}_k \cap \mathcal{C}$.
While, as we have seen previously, the feasible set in MPC-CBF formulation at each horizon step $k$ is $\mathcal{R}_k \cap \mathcal{S}_{\text{cbf},k}$.
Note that $\mathcal{S}_{\text{cbf},k}$ is a subset of $\mathcal{C}$. Therefore, the MPC-CBF formulation in \eqref{eq:mpc-cbf} has a smaller safe set than MPC-DC.

In the case the reachable set $\mathcal{R}_k$ is a proper subset of the safe set $\mathcal{C}$, then the distance constraints in \eqref{eq:mpc-distance-constraint} are not active in the optimization \eqref{eq:mpc-ftoc}, as shown in Fig.~\ref{fig:mpc-cbf-vs-mpc-dc-constraint-activation}a.
In other words, the distance constraints will not be active in the optimization \eqref{eq:mpc-ftoc} until the reachable set along the horizon intersects with the unsafe regions, i.e., the reachable set intersects the obstacles as shown in Fig.~\ref{fig:mpc-cbf-vs-mpc-dc-constraint-activation}b.
Using our MPC-CBF formulation, the CBF constraints in \eqref{eq:mpc-cbf-cbf} could always confine the reachable set $\mathcal{R}_k$ with an appropriate choice of $\gamma$ whenever the robot tends to approach the obstacles. In this case, even when the reachable set $\mathcal{R}_k$ is a proper subset of the safe set $\mathcal{C}$, the reachable set could still be constrained by its intersection with $\mathcal{S}_{\text{cbf},k}$, as shown in Fig.~\ref{fig:mpc-cbf-vs-mpc-dc-constraint-activation}a.

\begin{remark}
We discuss the constraint activation only in the cases when the robot is moving towards the obstacles, i.e., there exists a control input $\mathbf{u}$ such that $h(f(\mathbf{x}, \mathbf{u})) < h(\mathbf{x})$. When the robot is moving away from the obstacles, both distance constraints and CBF constraints are inactive, which is intuitive since the robot is always safe in this case.
\end{remark}
\begin{remark}
Our MPC-CBF formulation is safer than MPC-DC in the context of smaller set invariance, where we can see that $\mathcal{R}_k \cap \mathcal{S}_{\text{cbf},k}$ is a proper subset of $\mathcal{R}_k \cap \mathcal{C}$, as shown in Fig. \ref{fig:mpc-cbf-vs-mpc-dc-constraint-activation}.  
\end{remark}

\begin{figure*}
    \centering
    \begin{subfigure}[t]{0.19\linewidth}
        \centering
        \includegraphics[height = 0.99\linewidth]{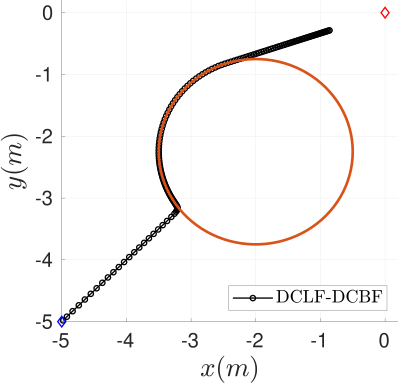}\\
        \caption{}
        \label{subfig:dclf-dcbf-avoidance}
    \end{subfigure}
    \begin{subfigure}[t]{0.19\linewidth}
        \centering
        \includegraphics[height = 0.95\linewidth]{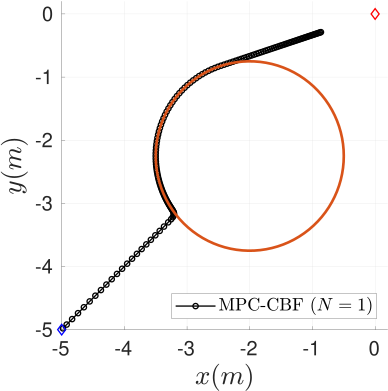}
        \caption{}
        \label{subfig:mpc-cbf-one-step-avoidance}
    \end{subfigure}
    \begin{subfigure}[t]{0.19\linewidth}
        \centering
        \includegraphics[height = 0.95\linewidth]{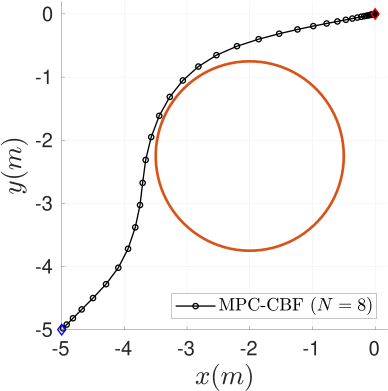}
        \caption{}
        \label{subfig:mpc-cbf-several-steps-avoidance}
    \end{subfigure}
    \begin{subfigure}[t]{0.19\linewidth}
        \centering
        \includegraphics[height = 0.95\linewidth]{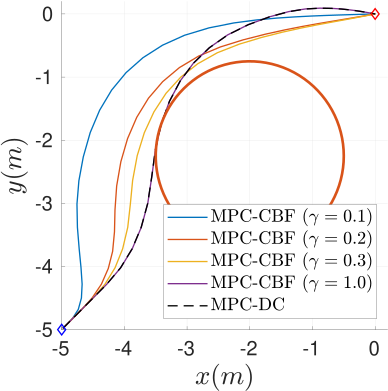}
        \caption{}
        \label{subfig:mpc-cbf-vs-mpc-dc-different-gamma}
    \end{subfigure}
    \begin{subfigure}[t]{0.19\linewidth}
        \centering
        \includegraphics[height = 0.95\linewidth]{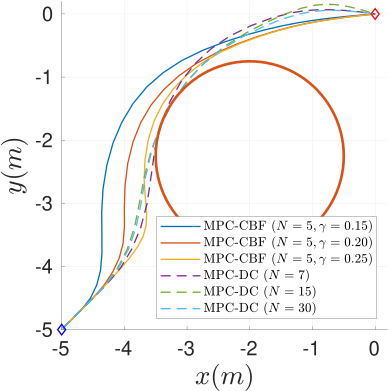}
        \caption{}
        \label{subfig:mpc-cbf-vs-mpc-dc-different-horizon}
    \end{subfigure}
    \caption{A 2D double integrator avoids an obstacle using different control designs. The obstacle is represented by a red circle and the start and target positions are located at $(-5,-5)$ and $(0, 0)$, labelled as blue and red diamonds, respectively. (a) a DCLF-DCBF controller; (b) a MPC-CBF controller with $N = 1$; (c) a MPC-CBF controller with $N = 8$ and $\gamma = 0.5$; (d) a MPC-DC controller with $N = 8$ and four MPC-CBF controllers with $N=8$ and different choices of $\gamma$; (e) three MPC-CBF controller with $N = 5$ and different values of $\gamma$ and three MPC-DC controllers with different values of horizon $N$. Notice that for $N = 5$, MPC-DC becomes infeasible when the state is close to the boundary of the obstacle, whose trajectory is therefore excluded from (e).}
    \label{fig:mpc-cbf-performance}
\end{figure*}

\begin{remark}
\label{remark:less-horizon}
In practice, we may need to use a smaller horizon for speeding up the optimization.
To achieve a similar performance compared to the MPC-DC formulation in \eqref{eq:mpc-ftoc}, we could apply smaller $\gamma$ in our MPC-CBF control design in \eqref{eq:mpc-cbf} and use a smaller horizon.
This could help us to reduce the complexity of the optimization for obstacle avoidance, as will be shown in Fig. \ref{subfig:mpc-cbf-vs-mpc-dc-different-horizon}.
\end{remark}

%% file: example.tex
\section{Examples}
\label{sec:example}

Having presented the MPC-CBF control design, we now numerically validate it using a 2D double integrator for obstacle avoidance and analyze its properties. We also apply this controller to a competitive car racing problem. We use the solver IPOPT~\cite{wachter2006implementation} in MPC-CBF/MPC-DC/DCLF-DCBF problems \cite{biegler2009large}.

\subsection{2D Double Integrator for Obstacle Avoidance}
\label{subsec:double-integrator}

Consider a linear discrete-time 2D double integrator
\begin{equation}
    \mathbf{x}_{k+1} = A \mathbf{x}_k + B \mathbf{u}_k,
    \label{eq:linear_dynamics}
\end{equation}
where the sampling time $\Delta t$ is set as $0.2s$.

A MPC-CBF is designed as in \eqref{eq:mpc-cbf} and \eqref{eq:mpc-cbf-law} for 2D double integrator to avoid an obstacle, where the stage cost and and terminal cost are
\begin{equation}
    q(\mathbf{x}_k, \mathbf{u}_k) = \mathbf{x}_k' Q \mathbf{x}_k + \mathbf{u}_k' R \mathbf{u}_k, \ p(\mathbf{x}_N) = \mathbf{x}_N' P \mathbf{x}_N, \label{eq:expression-cost}
\end{equation}
where $Q = 10\cdot\mathcal{I}_4$, $R = \mathcal{I}_2$ and $P = 100\cdot\mathcal{I}_4$. The system is subject to state constraint $\mathcal{X}$ and input constraints $\mathcal{U}$,
\noindent
\begin{gather*}
    \mathcal{X} = \{\mathbf{x}_k \in \mathbb{R}^n: \mathbf{x}_{\min} \leq \mathbf{x}_k \leq \mathbf{x}_{\max}\}, \\
    \mathcal{U} = \{\mathbf{u}_k \in \mathbb{R}^m: \mathbf{u}_{\min} \leq \mathbf{u}_k \leq \mathbf{u}_{\max}\}.
\end{gather*}
\noindent
The lower and upper bounds are
\noindent
\begin{equation*}
    \mathbf{x}_{\max}, \mathbf{x}_{\min} = \pm 5\cdot\mathcal{I}_{4\times1}, \ 
    \mathbf{u}_{\max}, \mathbf{u}_{\min} = \pm \mathcal{I}_{2\times1}.
\end{equation*}
\noindent
For discrete-time control barrier function constraint \eqref{eq:mpc-cbf-cbf}, we choose a quadratic barrier function for obstacle avoidance
\begin{equation}
    h_k = (\mathbf{x}_k(1) - x_{obs})^2 + (\mathbf{x}_k(2) - y_{obs})^2 - r_{obs}^2, \label{constraint-cbf-h}
\end{equation}
where $x_{obs}$, $y_{obs}$, and $r_{obs}$ describe x/y-coordinate and radius of the obstacle with $x_{obs} = -2m$, $y_{obs} = -2.25m$ and $r_{obs} = 1.5m$, shown as a red circle in Fig. \ref{fig:mpc-cbf-performance}.
The start and target positions are $(-5,-5)$ and $(0, 0)$, which are labelled as blue and red diamonds in Fig. \ref{fig:mpc-cbf-performance}, respectively.

\subsubsection{Comparison with DCLF-DCBF}
In order to compare the performance between our proposed MPC-CBF and DCLF-DCBF, we develop a DCLF-DCBF controller for the same obstacle avoidance task, which is based on \eqref{eq:CLF-CBF-discrete}.

For the design of DCLF-DCBF, we use $R$ in \eqref{eq:expression-cost} as $H$ in \eqref{eq:CLF-CBF-discrete} to ensure that we have the same penalty on inputs.
The discrete-time CBF constraints of DCLF-DCBF are the same as the ones used in MPC-CBF.
Since the terminal cost $p$ in the model predictive control serves as a control Lyapunov function, we choose $p$ in MPC-CBF example as the control Lyapunov function that is used to construct the discrete-time CLF constraints in \eqref{eq:CLF-CBF-discrete}.
Based on these choices, it is fair to compare a DCLF-DCBF controller with a MPC-CBF controller.

The simulation result of MPC-CBF and DCLF-DCBF comparison is shown in Fig. \ref{subfig:dclf-dcbf-avoidance}, \ref{subfig:mpc-cbf-one-step-avoidance} and \ref{subfig:mpc-cbf-several-steps-avoidance}. The trajectory is denoted in black line with small circles representing each step.
The trajectory of DCLF-DCBF controller with $\gamma = 0.4$ is presented in Fig. \ref{subfig:dclf-dcbf-avoidance}, where the system does not start to avoid the obstacle until it is close to it.
Fig. \ref{subfig:mpc-cbf-one-step-avoidance} shows the trajectory for MPC-CBF controller with horizon $N = 1$ and $\gamma$ = 0.4. This trajectory is similar to that of DCLF-DCBF controller.
Based on our analysis in Sec. \ref{subsec:relation-dclf-dcbf}, the performances of DCLF-DCBF and MPC-CBF with $N = 1$ are almost the same, which is validated in this simulation.
Fig. \ref{subfig:mpc-cbf-several-steps-avoidance} shows the trajectory of MPC-CBF controller with horizon $N = 8$ and $\gamma = 0.4$.
We can see that this controller can drive the system to avoid the obstacle earlier than the DCLF-DCBF controller. Also, among these three controllers, it is the only one that can reach the goal position in the limited simulation time.

\subsubsection{Comparison with MPC-DC}
A MPC-DC controller is developed based on \eqref{eq:mpc-ftoc} using the same parameters as MPC-CBF presented before except for the discrete-time CBF constraint, which is replaced by a Euclidean norm distance constraint $g$ shown in \eqref{eq:mpc-distance-constraint}. The function $g$ has the same expression as $h$, defined in \eqref{constraint-cbf-h}.


\begin{table}
    \centering
    \begin{tabular}{|c|c|c|c|c|c|c|} \hline
       controller & status & $N$ & $\gamma$ & mean/std (s) & min dist & cost \\ \hline
       MPC-CBF & solved & $5$ & $0.1$ & 0.028$\pm$0.012 & 1.483 & 7.620 \\
       MPC-CBF & solved & $5$ & $0.2$ & 0.028$\pm$0.011 & 0.791 & 7.464 \\
       MPC-CBF & solved & $5$ & $0.3$ & 0.028$\pm$0.011 & 0.441 & 8.314 \\
       MPC-CBF & solved & $5$ & $0.4$ & 0.028$\pm$0.011 & 0.288 & 8.292 \\
       MPC-CBF & solved & $5$ & $0.5$ & 0.028$\pm$0.010 & 0.110 & 8.813 \\ \hline
       controller & status & \multicolumn{2}{c|}{$N$} & mean/std (s) & min dist  & cost \\ \hline
       MPC-DC & infeas. & \multicolumn{2}{c|}{5} & NaN & NaN & NaN \\
       MPC-DC & solved & \multicolumn{2}{c|}{7} & 0.033$\pm$0.013 & 0.000 & 9.102 \\
       MPC-DC & solved & \multicolumn{2}{c|}{15} & 0.048$\pm$0.016 & 0.000 & 8.537 \\
       MPC-DC & solved & \multicolumn{2}{c|}{30} & 0.062$\pm$0.031 & 0.000 & 8.528 \\
       \hline
    \end{tabular}
    \caption{MPC-DC and MPC-CBF benchmark in terms of prediction horizon, computational time, minimal distance with respect to obstacle and cost integral. We need larger horizon for MPC-DC, otherwise the system only has noticeable obstacle avoidance behavior when it's close to the obstacles.}
    \label{tab:benchmark-computational-time}
\end{table}

Fig.~\ref{subfig:mpc-cbf-vs-mpc-dc-different-gamma} and \ref{subfig:mpc-cbf-vs-mpc-dc-different-horizon} show the simulation result of the comparison between MPC-CBF and MPC-DC.
In Fig. \ref{subfig:mpc-cbf-vs-mpc-dc-different-gamma}, MPC-CBF controllers with $\gamma$ = 0.1, 0.2, 0.3, 1.0 are described in blue, orange, yellow and purple lines, respectively.
The trajectory of MPC-DC controller is shown in black dashed line.
As $\gamma$ decreases, the system starts to avoid the obstacle earlier, which means a smaller safe set as analyzed in Sec. \ref{subsec:feasibility}, while on the other hand the trajectory of MPC-DC is the closest to the obstacle.
We also notice that the trajectories of MPC-DC and MPC-CBF with $\gamma=1$ are almost the same, which validates our analysis in Sec. \ref{subsec:relation-mpc-dc}.

The trajectories of MPC-CBF controllers with $N = 5$ and different choices of $\gamma$ and MPC-DC controllers with different values of horizon $N$ are shown in Fig. \ref{subfig:mpc-cbf-vs-mpc-dc-different-horizon}.
We notice that MPC-CBF controller with smaller $\gamma$ and MPC-DC with larger horizon $N$ can make the system avoid obstacles earlier.
This verifies our analysis in Sec. \ref{subsec:feasibility}, since smaller $\gamma$ in MPC-CBF and larger horizon $N$ in MPC-DC make the trajectory deviate from obstacles earlier.
We also observe that even with an extremely large horizon $N$, e.g. $N=30$, the system only has noticeable obstacle avoidance behavior when it is close to obstacles.
In contrary, a relatively small $\gamma$ is able to make the system avoid obstacles even when far away from obstacles.

In Fig. \ref{subfig:mpc-cbf-vs-mpc-dc-different-horizon}, MPC-CBF with $N = 5$ and $\gamma = 0.25$ starts to turn to avoid the obstacle with a similar behavior as MPC-DC with $N = 7$.
This property is discussed in Remark \ref{remark:less-horizon}.
Since discrete-time CBF enforces the invariance of safe set, it allows a smaller $N$ for MPC-CBF with a smaller $\gamma$ to achieve a comparable performance as MPC-DC with a larger $N$.

In Table \ref{tab:benchmark-computational-time}, we benchmark the MPC-CBF and MPC-DC in terms of prediction horizon, computation time, minimal distance to the obstacle and cost integral $\sum_k \mathbf{u}_k^T \mathbf{u}_k \Delta t$ over the trajectory.
We can observe that less prediction horizon of MPC-CBF leads to less computational time.
MPC-DC always reaches to the boundary of the obstacle, however, MPC-CBF could automatically set a safety margin depending on different choices of $\gamma$.
For this specific scenario, the cost integral over the trajectory of MPC-CBF is generally less than the one of MPC-DC.

\subsection{Competitive Car Racing}
\label{subsec:car-racing}
We have evaluated the MPC-CBF design using a 2D double integrator and compared its performance with DCLF-DCBF and MPC-DC. 
We proceed to implement MPC-CBF in a more complex scenario: competitive car racing.
In some previous car racing control work \cite{liniger2015optimization, verschueren2014towards}, they only consider static obstacles on the track while we deal with dynamic obstacles such as other cars using MPC-CBF.

\subsubsection{Vehicle Model}
\begin{figure}[h]
    \centering
    \includegraphics[width=0.8\linewidth]{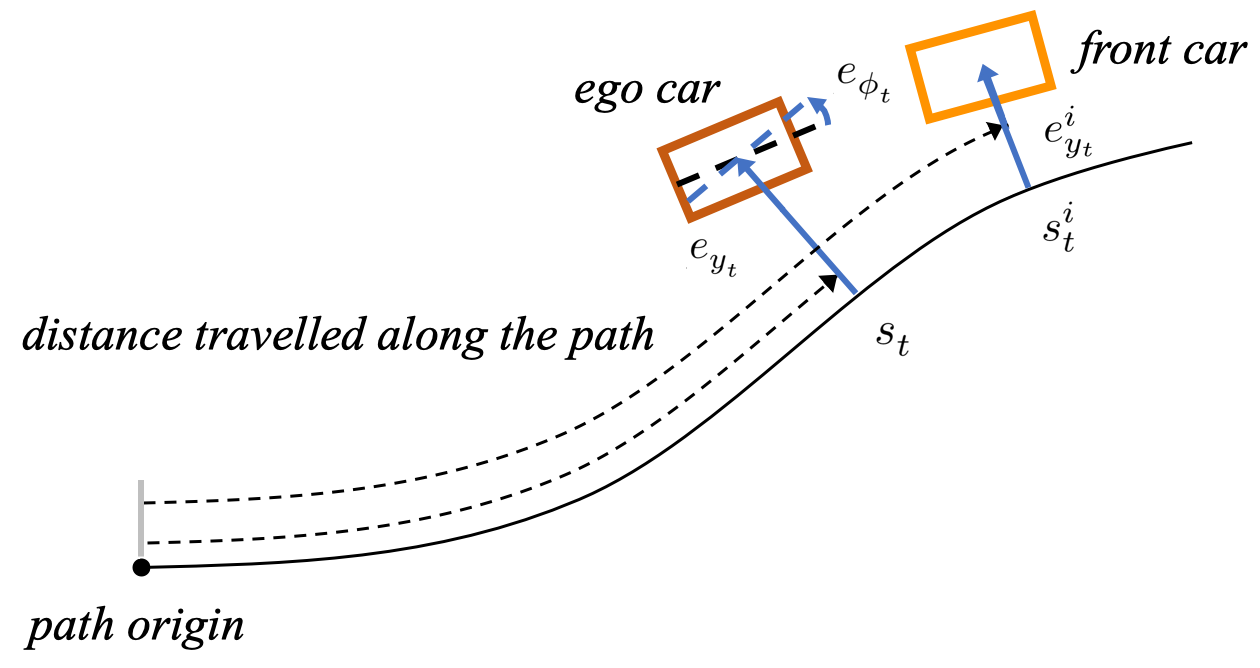}
    \caption{Representation of the ego car and front car in the curvilinear coordinate frame.}
    \label{fig:curvilinear}
\end{figure}

We use curvilinear coordinates to describe vehicle states of the ego and other cars in a racing competition.
In this paper, we use the nonlinear lateral vehicle dynamics model in \cite[Ch. 2]{rajamani2011vehicle} for system dynamics
\begin{equation}
    \mathbf{x}_{t+1} = f(\mathbf{x}_t, \mathbf{u}_t), \label{eq:lmpc_dynamics}
\end{equation}
where $x_t$ and $u_t$ represent the state and input of the vehicle at time step $t$ and their definitions are as follow
\begin{equation}
    \mathbf{x}_t = [v_{x_t}, \ v_{y_t}, \ \phi_{t}, \ e_{\phi_t}, \ s_t, \ e_{y_t}]^T, \ \mathbf{u}_t = [a_t, \ \delta_t]^T,
\end{equation}
where $s_t$ represents the curvilinear distance travelled along the centerline of the track, $e_{y_t}$ and $e_{\phi_t}$ are
the deviation distance and heading angle error between vehicle and path. $v_{x_t}$, $v_{y_t}$, $\phi_{t}$ are the vehicle longitudinal velocity, lateral velocity and yaw rate in the curvilinear coordinates, respectively.
A representation of the state in the curvilinear coordinate is shown in Fig. \ref{fig:curvilinear}.
The inputs are longitudinal acceleration $a_t$ and steering angle $\delta_t$. In the car racing, we assume to have $K$ racing cars competing with the ego car, and we use the superscript $i$ to distinguish the $i$-th ($i = 1,2,...,K$) competiting vehicle from the ego one, shown in Fig. \ref{fig:curvilinear}.
A detailed implementation is discussed in the Appendix \ref{appendix:car-racing}.

\subsubsection{Control Design}
A MPC-CBF is developed for this competitive car racing example using \eqref{eq:mpc-cbf}.
The stage cost function is designed as follows
\begin{equation}
\begin{split}
    q(\mathbf{x}_{t+k|t},\mathbf{u}_{t+k|t}) = & (\mathbf{x}_{t+k|t} - \mathbf{x}_r)^T Q (\mathbf{x}_{t+k|t} - \mathbf{x}_r) \\
    & + {\mathbf{u}_{t+k|t}^T} R {\mathbf{u}_{t+k|t}},
\end{split}
\label{eq:cost_lmpc_over_horizon}
\end{equation}
where $\mathbf{x}_r = (v_t, 0, 0, 0, 0, 0)$, $Q = diag(10, 0, 0, 0, 0, 10)$ and $R = diag(1, 1)$. This cost function allows the ego car to track the centerline with a target speed $v_t$ while minimizing the tracking error from the centerline.

The motion of overtaking other racing cars is considered as CBF constraints in \eqref{eq:mpc-cbf-cbf}. At time step $t$, each CBF $h^i_t$ represents the safety criterion between ego car at $(s_t, e_{y_t})$ and $i$-th other racing car at $(s^i_t, e^i_{y_t})$, described in the curvilinear coordinates, shown in Fig. \ref{fig:curvilinear}.
We choose CBF in a quartic form as follows
\begin{equation}
    \label{eq:lmpc_cbf_eq}
    h^i_t = \dfrac{(s_t - s^i_t)^4}{(2l_1)^4} + \dfrac{(e_{y_t} - e^i_{y_t})^4}{(2l_2)^4} - 1,
\end{equation}
where we assume all racing cars including ego car hold the shape of rectangle with a length as $2l_1$ and width as $2l_2$. 
Notice that we assume that we have perfect estimation about $(s_t, e_{y_t})$ and $(s^i_t, e^i_{y_t})$.

\begin{figure} 
    \centering
    \includegraphics[width=1\linewidth]{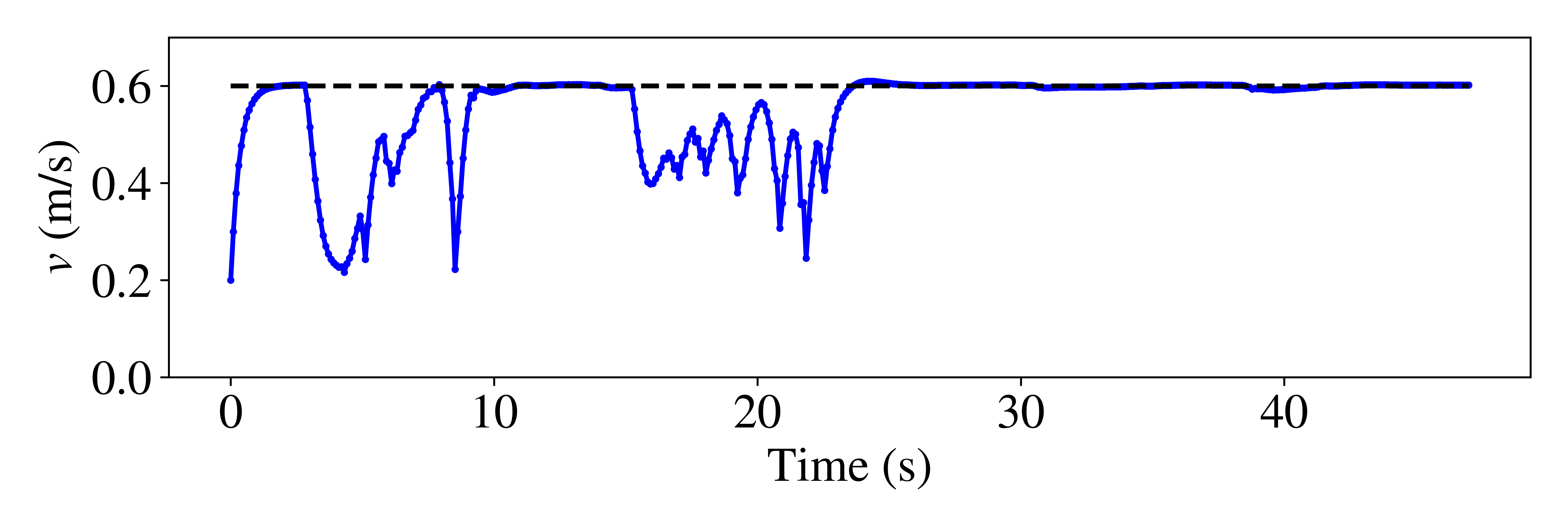}
    \caption{Speed profile during the car racing competition in one lap of the simulation. The dashed black line shows the desired speed $v_t = 0.6$m/s.}
    \label{fig:mpc-cbf-car-racing-speed-norm}
    \vspace{-5mm}
\end{figure}

\subsubsection{Simulation \& Results}
During the competition, we expect ego car to track the centerline with a target speed $v_t = 0.6$m/s.
MPC-CBF with a horizon $N$ = 12 updates at $10$ Hz.
The system dynamics is simulated at $1000$ Hz and the controller sampling time is $0.1s$.
Our ego vehicle is simulated with the nonlinear lateral vehicle dynamics model and we use the linearized dynamics along the centerline to formulate our control design.
In the simulation, we deploy several racing cars to compete with ego car. 
In order to better illustrate results, a snapshot of overtaking motion with a zoom-in view is shown in Fig. \ref{fig:cover}. Ego car begins with an initial speed $v_0 = 0.2$ m/s at the origin of the centerline and two other racing cars start in front of the ego car. Two cars are simulated to move at $0.2$ m/s while keeping a constant distance deviation $e^i_{y}$ from the centerline, where $e^1_{y} = 0.1$ m and $e^2_{y} = -0.1$ m. Fig. \ref{fig:cover} demonstrates that the MPC-CBF allows ego car to safely race and overtake other cars in both left and right directions.
Fig. \ref{fig:mpc-cbf-car-racing-speed-norm} shows the speed profile, where the dashed black line shows the desired speed.
We can see that ego car always tries to catch up to the target speed during the competitive car racing. 
In Fig. \ref{fig:cover}, we observe two motions of overtaking front racing cars. 
Since two racing vehicles hold opposite distance deviations from the centerline, ego car overtakes them with right and left turns respectively.

%% file: conclusion.tex
\vspace{-1mm}
\section{Conclusion}
\label{sec:conclusion}
A safety-critical model predictive control design is proposed in this paper, where discrete-time control barrier function constraints are used in a receding horizon fashion to ensure safety.
We present an analysis of its stability and feasibility, and describe its relation with MPC-DC and DCLF-DCBF.
To verify our analysis, we use a 2D double integrator for obstacle avoidance, where we can see that MPC-CBF outperforms both MPC-DC and DCLF-DCBF.
The proposed control logic is also applied to a more complex scenario: competitive car racing, where our ego car can race and safely overtake other racing cars.


\vspace{-1mm}
\section*{Acknowledgement}
We thank Ugo Rosolia for his insightful discussions.

%% file: appendix.tex
\appendix
\subsection{Car Racing Implementation}
\label{appendix:car-racing}
As mentioned in Sec. \ref{sec:example}, the vehicle dynamics is described by the kinematic model using the curvilinear coordinates under the Frenet reference frame. The vehicle discrete-time model is Euler discretized at time step $t$, shown as follow
\begin{align*}
    v_{x_{t+1}} &= v_{x_t} + dt \left( a_t - \dfrac{1}{m} F_{yf_t} \sin(\delta_t) + \phi_t v_{y_t} \right), \\
    v_{y_{t+1}} &= v_{x_t} + dt \left( \dfrac{1}{m} (F_{yf_t} \cos(\delta_t) + F_{yr_t}) - \phi_t v_{x_t} \right), \\
    \phi_{t+1} &= \phi_t + dt \left( \dfrac{1}{I_z} (l_f F_{yf_t} \cos(\delta_t)) - l_r F_{yr_t} \right), \\
    e_{\psi_{t+1}} &= e_{\psi_t} + dt \left(\phi_t - \dfrac{v_{x_t} \cos(e_{\psi_t}) - v_{y_t} \sin(e_{\psi_t}) }{1 - \kappa(s_t) e_{y_t}} \kappa(s_t) \right), \\
    s_{t+1} &= s_t + dt \left(\dfrac{v_{x_t} \cos(e_{\psi_t}) - v_{y_t} \sin(e_{\psi_t})}{1 - \kappa(s_t) e_{y_t}}\right), \\
    e_{y_{t+1}} &= e_{y_t} + dt \left( v_{x_t} \sin(e_{\psi_t}) + v_{y_t} \cos(e_{\psi_t}) \right),
\end{align*}
where $\kappa(s_t)$ represents the curvature and $F_{yf_t}$, $F_{yr_t}$ describe the lateral force at front and rear tire at time step $t$
\begin{align*}
    F_{yf_t} &= 2 D_f \sin(C_f \arctan(B_f \alpha_{f_t})), \\
    F_{yr_t} &= 2 D_r \sin(C_r \sin(C_r \arctan(B_r \alpha_{r_t}))),
\end{align*}
and $\alpha_{f_t}$ and $\alpha_{r_t}$ are the tire angles, holding the relation with respect to the system states and inputs as follows
\begin{align*}
    \alpha_{f_t} &= \delta_t - \arctan(\dfrac{v_{y_t} + l_f \phi_t}{v_{x_t}} ), \\
    \alpha_{r_t} &= -\arctan(\dfrac{v_{y_t} - l_f \phi_t}{v_{x_t}}).
\end{align*}
In the equations above, $l_f$, $l_r$, $B_f$, $B_r$, $C_f$, $C_r$, $D_f$, $D_r$, $m$, $I_z$ represent the vehicle parameters.

We note that this kinematic model is highly nonlinear and this model cannot be applied to formulate system dynamics constraints in \eqref{eq:mpc-cbf-dynamics}.
Instead, we follow the data-driven approach proposed in \cite{rosolia2019learning}. We firstly simulate the vehicle system to track the centerline by using a PID controller
\begin{align*}
    \delta_t &= -k_1 e_{y_t} - k_2 e_{\psi_t}, \\
    a_t &= k_3 (v_d - v_{x_t}),
\end{align*}
where $v_d$ could be any positive user-defined target speed.
This PID controller allows the vehicle to track the centerline of the track with a target speed.

Then, the linearized time-invariant dynamics is shown as below
\begin{equation}
    \label{eq:linearized-vehicle-dynamics}
    \mathbf{x}_{t+1} = A \mathbf{x}_t + B \mathbf{u}_t,
\end{equation}
where $A$ and $B$ are calculated with a regression-based approach \cite{rosolia2019learning} over the trajectory we simulated using a PID controller. This process allows us to approximate the highly nonlinear dynamics with the time-invariant linearized one in \eqref{eq:linearized-vehicle-dynamics}.
The whole process is purely data-driven without a prior knowledge of the system parameters.
This linearized dynamics \eqref{eq:linearized-vehicle-dynamics} is used in CBF constraints \eqref{eq:mpc-cbf-dynamics}, which reduces the burden of the computational complexity for the numerical simulation.